\documentclass{emulateapj} 

\slugcomment{ApJ in press}

\shorttitle{WISE QSOs at $z>6$}
\shortauthors{Blain et al.}

\begin{document}


\title{WISE detections of known QSOs at redshifts greater than six}


\author{Andrew W. Blain,\altaffilmark{1} Roberto Assef,\altaffilmark{2,3,4} Daniel Stern,\altaffilmark{2} Chao-Wei Tsai,\altaffilmark{2,3} Peter Eisenhardt,\altaffilmark{2} Carrie Bridge,\altaffilmark{5} Dominic Benford,\altaffilmark{6} 
Tom Jarrett,\altaffilmark{7} Roc Cutri,\altaffilmark{8} Sara Petty,\altaffilmark{9} Jingwen Wu,\altaffilmark{10} and Edward L. Wright\altaffilmark{10}}
\email{ab520@le.ac.uk}


\altaffiltext{1}{Physics \& Astronomy, University of Leicester, 1 University Road, Leicester, LE1 7RH, UK}
\altaffiltext{2}{Jet Propulsion Laboratory, California Institute of Technology, Pasadena, CA\,91109, USA}
\altaffiltext{3}{NASA Postdoctoral Program (NPP).}
\altaffiltext{4}{N\'ucleo de Astronom'a de la Facultad de Ingenier'a, Universidad Diego Portales, Av. Ej\'ercito 441, Santiago, Chile}
\altaffiltext{5}{California Institute of Technology, 249-17, Pasadena, CA\,91125, USA}
\altaffiltext{6}{Goddard Space Flight Center, Greenbelt, MD\,20771, USA}
\altaffiltext{7}{Astronomy Department, University of Cape Town, South Africa}
\altaffiltext{8}{Infrared Processing and Analysis Center, California Institute of Technology, MS100-22, Pasadena, CA\,91125, USA}
\altaffiltext{9}{Virginia Tech, Department of Physics MC 0435, 910 Drillfield Drive, Blacksburg, VA 24061, USA}
\altaffiltext{10}{Division of Astronomy \& Astrophysics, University of California, Los Angeles, Physics and Astronomy Building,
	430 Portola Plaza,
        Los Angeles, CA\,90095-1547, USA}



\begin{abstract}
We present WISE All-Sky mid-infrared (IR) survey detections of 55\% (17/31) of the known QSOs at $z>6$ 
from a range of 
surveys: the SDSS, the CFHT-LS, FIRST, {\it Spitzer} and 
UKIDSS. 
The WISE catalog thus provides a substantial increase in the quantity of IR data available for these 
sources: 17 are detected in the WISE W1 (3.4\,$\mu$m) band, 16 in W2 (4.6\,$\mu$m), 3 in W3 (12\,$\mu$m) and 0 in W4
(22\,$\mu$m). This is particularly important with {\it Spitzer} in its warm-mission phase and no faint follow-up capability 
at wavelengths longwards of 5\,$\mu$m until the launch of {\it JWST}. 
WISE thus provides a useful tool for understanding QSOs found in forthcoming large-area optical/IR sky surveys, using 
PanSTARRS, SkyMapper, VISTA, DES and LSST. The rest-UV properties of the WISE-detected and the WISE-non-detected samples differ: the detections have brighter 
$i$/$z$-band magnitudes and redder rest-UV colors. This suggests that a more aggressive hunt for very-high-redshift 
QSOs, by combining WISE W1 and W2 data 
with red observed optical colors 
could be 
effective at least for a subset of dusty candidate QSOs. 
Stacking the WISE images of the 
WISE-non-detected QSOs indicates that they are on average significantly fainter than the WISE-detected examples, 
and are thus not narrowly missing detection in the WISE catalog. 
The WISE-catalog detection of three of our sample in the 
W3 
band indicates that their mid-IR flux can be 
detected individually, although there is no stacked W3 detection of sources detected in W1 but  not W3.
Stacking analyses of WISE data for large AGN samples will be a useful tool, 
and high-redshift QSOs of all types will be easy targets for {\it JWST}. 
\end{abstract}


%



\section{Introduction}

At the highest redshifts, QSOs provide excellent tools for probing the process and end of reionization, and 
provide signposts to some of the most active locations and phases in the process of galaxy evolution. 
A variety of surveys have been used to select active galaxies at high redshifts on the 
grounds of their optical and near-infrared (IR) colors. The optical Sloan Digital Sky Survey (SDSS; Abazajian et al. 2009) 
has mapped one quarter of the sky, from which the high-redshift QSOs have been selected and studied by Fan et al. (2001, 
2003, 2004, 2006) and Jiang et al. (2008). A deeper, narrower optical--IR QSO survey has also been conducted by the 
Canada--France--Hawaii Telescope (CFHT) Legacy Survey (e.g., Willott et al. 2007, 2010b). Combining near-infrared imaging data in 
the selection criteria, the UKIDSS survey currently holds the QSO redshift record at $z=7.086$ (Mortlock et al. 
2011). The southern 2dF and 6dF surveys (Croom et al. 2004; Heath Jones et al. 2009), based on digitized 
optical plate imaging and the 2MASS IR all-sky survey respectively, provide a 
substantial increase in the area surveyed for color-selected spectroscopically-confirmed QSOs, 
beyond the footprint of SDSS. One further optically-faint QSO was independently identified in the 8.5-deg$^2$ {\it Spitzer} Deep, 
Wide-Field Survey in Bo\"otes  by McGreer et al. (2006) and Stern et al. (2007). 
A current summary of references to $z>6$ QSOs can be found in Wang et al. (2011b). 
Deep observations using {\it Spitzer} have confirmed that at least some of the highest-redshift QSOs 
are detectable in the mid-IR  (Charmandaris et al. 2004; Hines et al.\ 2006, Jiang et al.\ 2006, Stern et al.\ 2007), with properties not dramatically different than lower-redshift examples (Elvis et al.\ 1994; Staguhn et al. 2005). Some high-redshift QSOs are certainly also very rich in gas and dust, and provide most of the significant detections of ultraluminous galaxies in molecular lines at the highest redshifts (Wang et al.\ 2011a; Maiolino et al. 2012). 
There appears to be significant variation in the amounts of both the hottest and the total dust present; however, see Jiang et al. (2010). Some relevant results of analyzing spectral energy distributions (SEDs) of high-redshift active galaxies, based on wideband UV-radio observations, including data from {\it Spitzer} and {\it Herschel}, are presented in Leipski \& Meisenheimer (2012); in particular, the long-wavelength far-IR and submm properties of high-redshift active galaxies are discussed by Leipski et al. (2013).

The recent spectroscopic confirmation by Mortlock et al. (2011) of the first bright near-IR-selected 
QSO at $z>7$ spurred a search 
for the properties of known QSOs at high redshifts in the WISE All-Sky survey (Wright et al. 2010). WISE has
mapped the whole sky in four IR bands, centered at 3.4, 4.6, 12 and 22\,$\mu$m (bands W1-W4, respectively), 
with the most relevant bands W1 and W2 reaching a typical 5-$\sigma$ sensitivity of 6.8 and 9.8\,$\mu$Jy, respectively, 
and an angular resolution of about 6\,arcsec. As a result of the polar orbit of WISE, depths vary across the sky: there is deeper coverage at the poles than at the equator. 

We compiled a list of all 31 published spectroscopically-confirmed QSOs 
at 
$z>6$, and searched the WISE All-Sky data release products catalog (14th March 2012; Cutri et al.\,2012)\footnote{http://wise2.ipac.caltech.edu/docs/release/allsky/expsup} to find 
detections and limits of these objects in the WISE bands from 3.4 to 22\,$\mu$m. The $\simeq 6$\,arcsec spatial resolution 
of WISE is approximately three times coarser than the 1.7\,arcsec available from 
{\it Spitzer}-IRAC at 3.6\,$\mu$m, and the integration time per point on the sky from WISE is typically only two 
minutes; however, the images have excellent astrometry and a search of the WISE database provide sensitive
mid-IR ($>5\,\mu$m) 
all-sky imaging, 
reaching many times deeper 
than existing surveys from {\it IRAS} and {\it Akari}. Seven of our targets have archival detections by {\it Spitzer} 
(Jiang et al.\ 2006; Stern et al. 2007), but most were published after the cryogenic phase of {\it Spitzer} was completed, and so 
WISE provides the only source of mid-IR information longwards of 5\,$\mu$m, on most of these targets until the launch of {\it JWST}. 
Most of the detections are made in WISE bands W1 and W2, which approximately match the IRAC [3.6] and [4.5] bands
that are still operating in the {\it Spitzer} warm mission. Seventeen 
WISE detections are presented in Table\,1, 14 in the main all-sky 
release source catalog with $>5$-$\sigma$ detections -- and three from the `Reject Table'\footnote{http://
wise2.ipac.caltech.edu/docs/release/allsky/expsup/ sec2\_4a.html} that contains $> 3.5$-$\sigma$ detections. This  
confirms that {\it Spitzer} can easily continue to detect the highest-redshift QSOs while the warm mission continues. 

High-redshift QSOs are typically selected on the grounds of their smooth, broad, power-law spectra
redwards of a strong redshifted Lyman-break feature, that at these redshifts lies between the $i$ and $z$ bands for $z\sim6$ and between the $z$ and $J$ bands for the $z=7.02$ UKIDSS QSO. 
The WISE bands alone provide little additional power to select 
these objects, owing to the relatively featureless continuum emission of AGN yielding similar colors across the WISE bands at 
all redshifts (see Fig.\, 1); however, when WISE detects high-redshift AGN, the measured fluxes 
provide a valuable measure of the rest-frame optical spectral energy distribution in the W1 and 
W2 bands, and signs of hot dust emission from the immediate environment of the AGN can be studied using 
the W3- and W4-band data. 

In Section\,2 we discuss the WISE properties of the $z>6$ QSOs, and discuss their multiwavelength properties in Section\,3. 
The quoted magnitudes 
are in their native system (Vega for WISE, AB for SDSS), unless otherwise stated (note Figs 5 \& 6).

\begin{figure}
\includegraphics[width=3.27in]{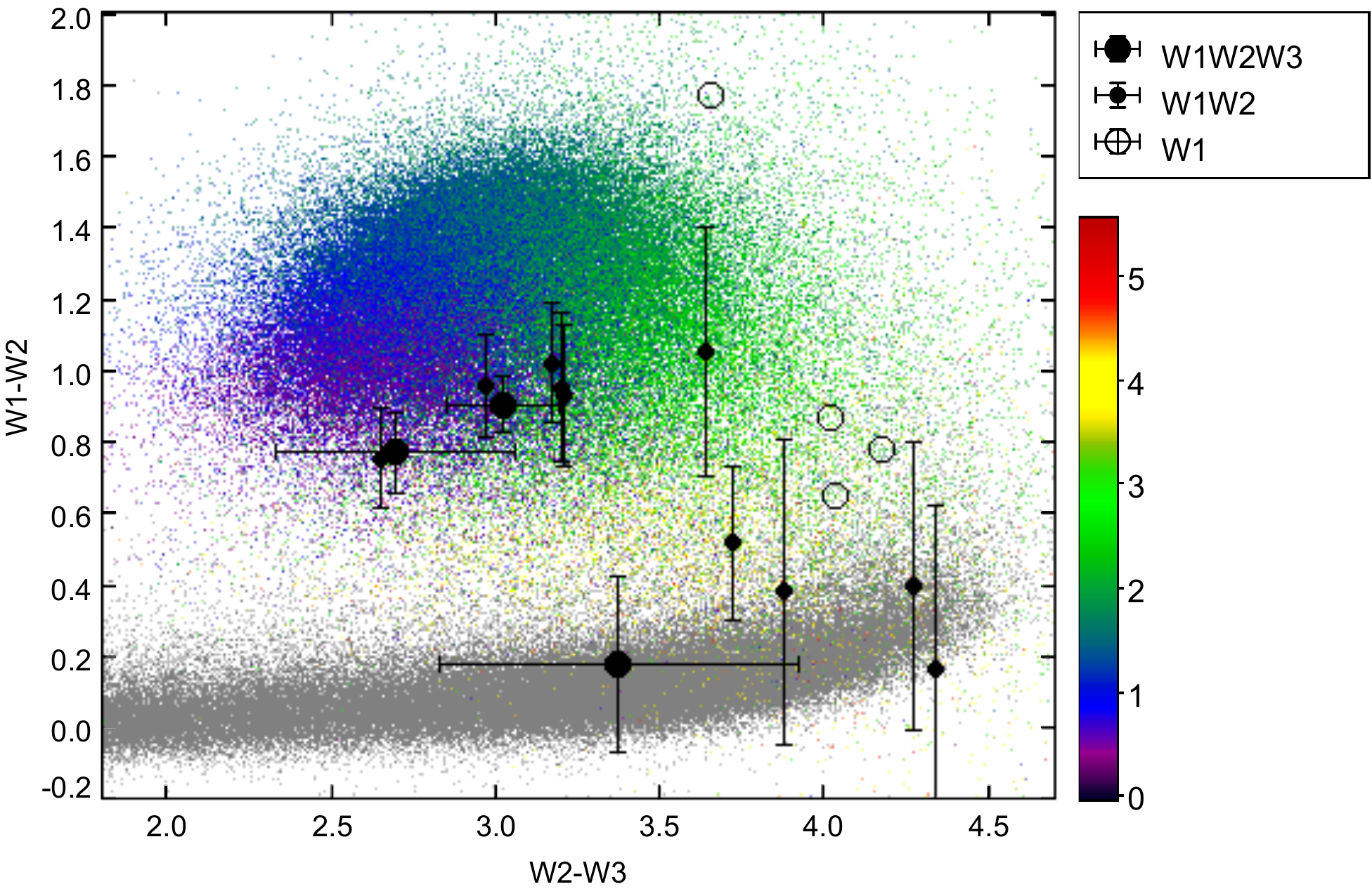}
\includegraphics[width=3.27in]{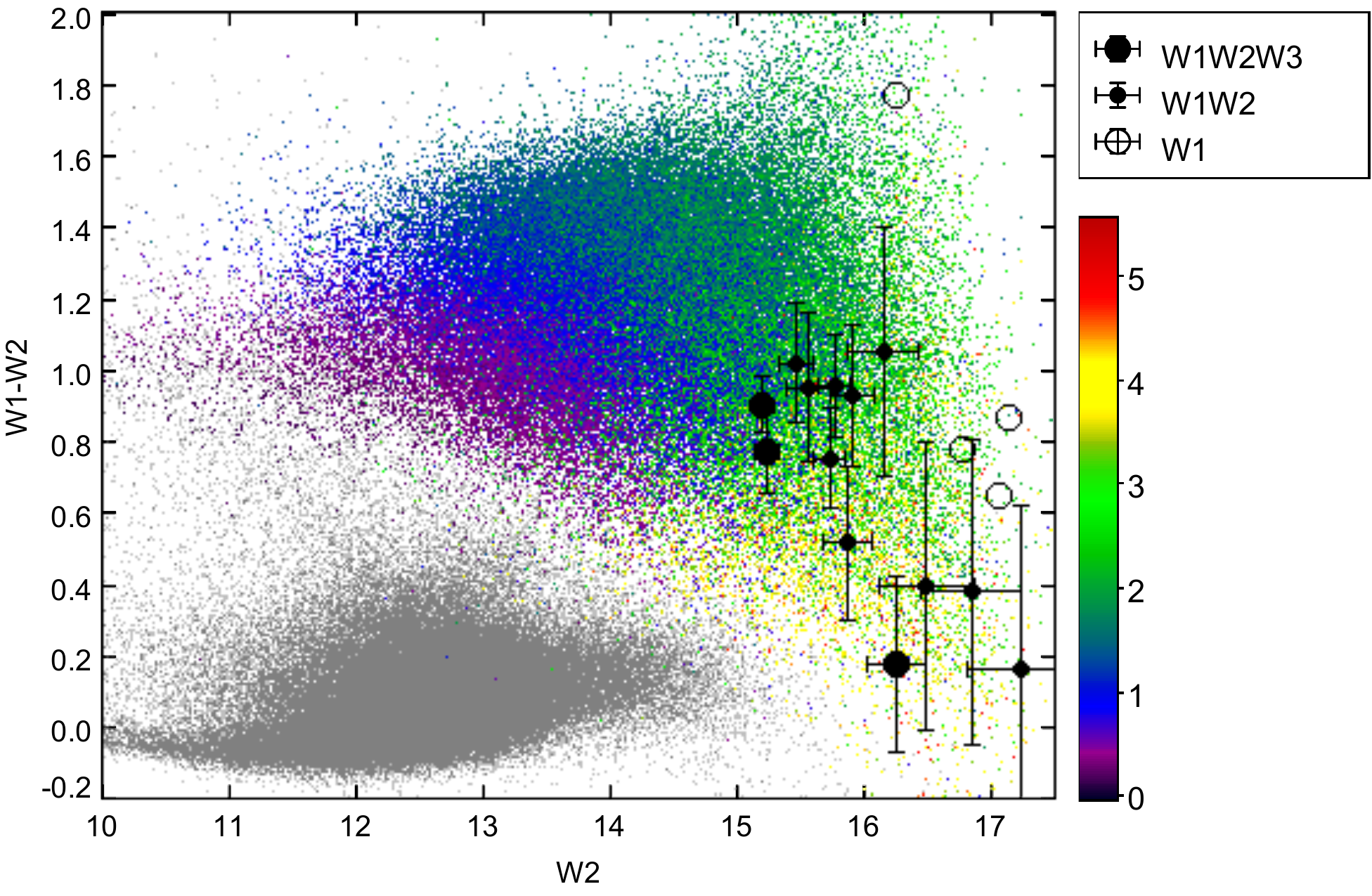}
\caption{WISE color-color and color-magnitude diagrams for $z>6$ QSOs (filled black circles), and a large 
sample of QSOs matched between SDSS and WISE (colored points) that are detected in all three relevant WISE bands. 
Large solid circles represent $z>6$ QSOs objects with detections in all three of the bands W1/W2/W3. Smaller solid circles represent two-band detections in W1/W2. Open circles represent detections in W1 
alone, and are included for completeness only. Where there is no error bar in the W1-W2 vs W2-W3 plot, the 
filled symbols represents an upper limit to the value of the W2-W3 color.  
The mid-IR properties of the SDSS QSO sample at $z<5.4$ (colored field; Schneider et al.\ 2007) were obtained 
from a cross match with the WISE All-sky Data Release catalog. 
Catalogued WISE objects within 3.5\,arcsec of the reported SDSS QSO survey positions were assumed to be correct 
identifications. The gray cloud of points represents a sample of low-redshift galaxies (see Wright et al. 2010 and Jarrett et al. 2011) 
to illustrate the underlying galaxy populations. 
}
\end{figure}

\section{WISE properties of QSOs identified at $z>6$} 

The list of 17 $z>6$ QSOs that have cataloged WISE detections 
are presented in Table\,1, along with 
their flux densities or limits from the WISE catalogs. 
Thirteen of the 14 all-sky release detections in W1 are also detected in the W2 band; a subset of three are also listed as 
detections in W3; none are detected in W4. All of the reported detections -- which are distinguished in the table by not having a null 
(...) entry for a noise estimate -- have a WISE pipeline signal-to-noise ratio 
value (wXsnr) greater than 3.5 for band X=1, 2 or 3. 
The WISE colors of these high-redshift QSOs are compared with a match of the WISE and the SDSS QSO catalogs 
(Schneider et al.\ 2007) in Fig.\,1. That match yielded WISE colors for 19,876 QSOs at $z<5.6$. 

\begin{figure}
\includegraphics[width=3.5in]{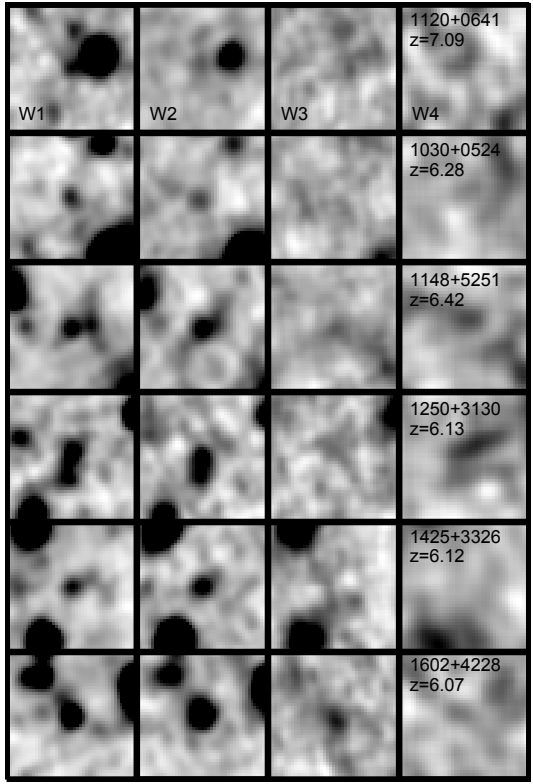}
\caption{Postage stamp images, 60-arcsec on a side, centered on the QSO position in WISE bands W1, W2, W3, W4 (from left to right along each row). From top to bottom we show a different QSO in each row (labelled in the last column with their redshifts): 
ULAS\,J1120+0641, 
SDSS J1030+0524, 
SDSS J1148+5251, 
QSO J1250+3130, 
QSO B1425+3326/SDWFS\,J1427+3312 
and SDSS J1602+4228. 
QSO J1250+3130 has a nearby companion, but far enough away from the WISE counterpart to ensure that the 
photometry remains accurate. The greyscale intensity is chosen to give an effective representation of the noise, and 
so bright sources are saturated black. 
}
\end{figure}

The WISE-detected high-redshift QSOs are all found at similar positions in the WISE W1-W2:W2-W3 color space, close to (3.0, 1.0), deep 
within the densest region of WISE colors for lower-redshift AGN (Fig.\,1). This highlights the difficulty of using WISE photometry 
alone to identify candidate high-redshift QSOs. Where there are only WISE limits, the regions where QSOs may lie covers 
almost the whole of this color space. 
Combining WISE and optical/near-IR data to highlight 
candidates is more promising as we discuss in Section\,3. 

The WISE depth of coverage varies across the sky, with depth increasing away from the ecliptic plane towards the poles.
The detectability of a target QSO appears 
at least in part to be related to this factor. The number of 
individual imaging frames taken by WISE that were included in the creation of the WISE Atlas Images for the 14 
detections in the All-Sky Release source catalog ranged from 11--48, with a median of 21 and an interquartile range of 13--28. These numbers are unchanged when the three Reject-Table QSOs are included. For the 14 WISE non detections, the number of individual WISE 8-s exposures that contributed to the catalog entry was 12--24, with a median of 15, and an interquartile range of 13--22. 
We discuss the results of stacking a total of 188 of these 8-s single-exposure WISE image frames for the 
10 non-detected sources 
below, and conclude that, based on their different WISE fluxes, this coverage effect alone is unlikely to 
account for the difference between the detections and non-detection of high-redshift QSOs by WISE.  
 
 The WISE Atlas Images corresponding to the high-redshift QSOs were inspected to ensure that the reported source was not affected 
 by image artifacts, or other problems. ULAS\,J1120+0641 is within 15\,arcsec of a bright star; however, the photometry is 
 uncontaminated. The images for the undetected $z>6$ QSOs were also inspected, and while 
 in some cases, as expected 
 net positive signals are found at the QSO position, they are all consistent with a non-detection, and thus are 
 correctly reported  as non-detections in the  pipeline-processed WISE catalog. 
Further data was taken in the W1 and 
W2 bands after the cryogen was exhausted, the `WISE Post-Cryo Survey Phase' of the mission. These data will be combined 
with the cryogenic mission data in the near future, resulting in an `All-WISE' catalog, with an anticipated noise level about 1.4 times deeper over about 70\,per cent of the sky, increasing the utility of WISE for detecting QSOs samples further.\footnote{NOTE ADDED IN PROOF:Ê
The enhanced AllWISE data release is scheduled for late 2013. The AllWISE catalogs are being compiled as this paperÊ
goes to press. Sixteen of the sources listed in Table 1 are in the new source tables. TheÊ
exception is SDSS J2054-0005, which was listed in the All-sky Reject Table. Searching the 31 high-redshiftÊ
targets in the new tables yields one additional target that satisfies the conditions for inclusion in theÊ
All-Sky Release Catalog and three additional targets that satisfy the conditions for being entered in the All-SkyÊ
release Reject Table, while two sources are promoted from the Reject Table to the Catalog, and one source is demoted from theÊCatalog to the Reject Table. Hence, the 14 targets listed in the All-Sky Release Catalog, and the 3 in the Reject Table inÊ
Table\,1 for the WISE All-Sky data release are likely to be changed to 16 and 4 respectively in the AllWISE Catalog,Ê
giving an increased catalogued fraction of 65 per cent for known QSOs at redshifts greater than six.}
 
 \begin{figure}
\epsscale{1.15}
\vskip 15pt
\plotone{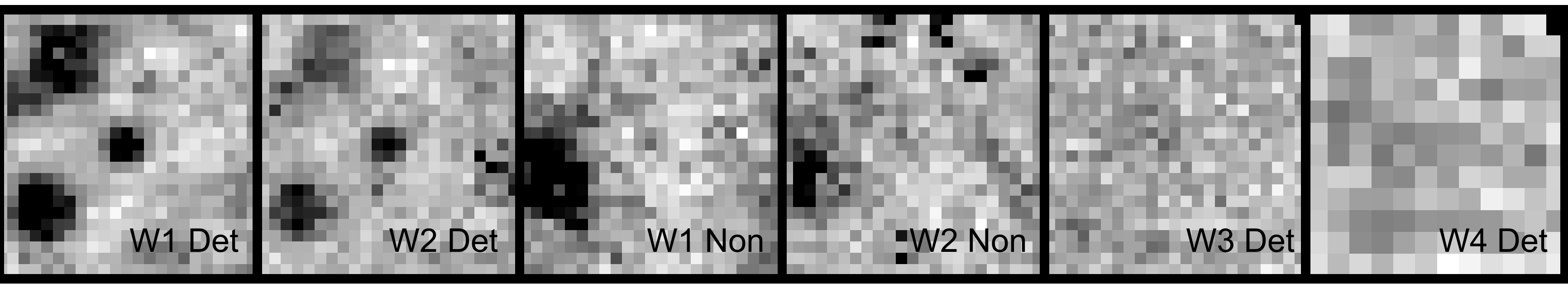}
\caption{The results of stacking various combinations of the WISE images of $z>6$ QSOs. From left to right, a stack of the W1 and W2 images for 16/17 of the W1-detected QSOs are shown as a comparison (W1 Det and W2 Det). Stacked W1 and W2 
images for 10/14 of the non-detected objects with no nearby bright sources 
are then shown (W1 Non and W2 Non). Finally W3 and W4 images are stacked from 13/14 W1-detected objects without 
W3 detections (W3 Det and W4 Det). 
Sources were excluded from the stacks on the grounds of a nearby brighter source that contributed light at the center of the stack; ULAS\,1120+0641 alone was excluded from the detected (Det) sources. 
The total number of 8-s-duration WISE single-exposure image frames combined in these stacks is 300, 300, 188, 188, 297 and 297 respectively.}
\end{figure}

Five QSOs -- SDSS\,J1030+0524, SDSS\,J1148+5251, QSO\,J1250+3130, QSO B1425+3326, and SDSS\,J1602+4228 -- have 
bright W1 and W2 detections. Their WISE images, along with the images of the 
highest-redshift target ULAS\,J1120+0641, are shown in 
Fig.\,2;\footnote{We also compare the observed-frame optical-mid-IR SEDs for four of these objects in Fig. 4. We exclude QSO B1425+3326 owing to concerns about a contaminating source, 
although its $i$-band to WISE colors are not unusual as compared with the other detections.} 
In the case of SDSS\,1030+0524, SDSS\,J1148+5251 and QSO B1425+3326 there are {\it Spitzer} measurements (Jiang et al. 2006; Stern et al.\ 2007) that are consistent with the WISE W1 and W2 flux densities reported in Table\,1. Three have a relatively large number of detections, whereas QSO\,J1250+3130 is a very typical source. 

In order to investigate the non-detections further, both for the W1 and W2 bands in the non-detected targets, and for the 
W3 and W4 bands in the detected targets, we stacked the WISE images from the QSO positions. In order to overcome the 
variations in the coverage of different targets, the whole ensemble of 8-s-duration single-exposure WISE images that 
overlapped the target positions were extracted from the WISE Image Atlas, and stacked with hot pixels and bright sources 
masked in order to probe the net flux density of the sample. This was done for all the targets without nearby bright objects, 
corresponding to 300 W1 frames, 300 W2 frames, 297 W3 frames and 297 W4 frames for 16 (of 17 in total) high-redshift QSOs with detections in the bluer WISE bands. Stacking the frames for the 10 sources uncontaminated by nearby brighter objects (of 14 in total) without WISE detections in W1 or W2 provided a total of 188 frames in each of these bands. 
These stacked images are shown in Fig.\,3, with 
the same grayscale display range as shown in Fig.\,2. Note that the pixel scale in the W4 images is twice that in the other 
three bands. There is reasonably a clear detection of the WISE-catalogued objects in the stack in both the W1 and W2 bands, at S/N ratios 
of 13.5 and 10.9 respectively. 
By comparison, the stack of the non-detected QSOs shows no sign of emission in the W1 or W2 bands (Fig.\,3), with 
a noise level in the stacked image that is 26 per cent greater, 
reflecting the lesser number of uncontaminated fields included in the stack of non-detected targets (188 vs 300).
 There are no indications of any net detections in the longer WISE bands, W3 and W4, for any subsamples, which is reasonable given the 
sensitivity and the modest number of stacked fields.

\begin{figure}
\epsscale{1.2}
\vskip -30pt
\plotone{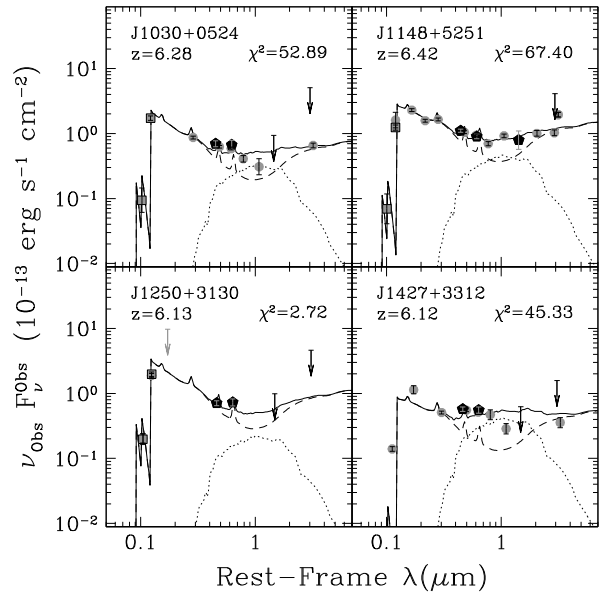}
\vskip -50pt
\caption{SEDs of four secure WISE 2/3-band detections from the $z>6$ QSO sample. 
A combination of  different SED component templates  (Assef et al.\ 2010) which were developed for data in the NDWFS 
B\"ootes field, and are adequate to describe the WISE data are 
also shown, for an AGN (solid line) and an evolved stellar population (dotted line). The addition of other template components, for example for a star-forming galaxy (Sbc/Im) component 
does not improve the fits. The WISE data is highlighted by filled black pentagons, and upper limits at the longer wavelengths. SDSS data is represented by gray squares, and {\it Spitzer} and other sources as gray circles. 
J1250+3130 provides a good example of how data from WISE can provide a valuable insight into the relative strengths of different SED components from a typical WISE-detected example without {\it Spitzer} data. The sources of the data are referenced in Table\,1. In all four cases, the WISE data and limits in W1 and W2 clearly limit the possible contribution from an evolved stellar population, while the information from W3 and W4 provides a valuable limit to the possible normalization and/or slope of the power-law mid-IR AGN component in the event that Spitzer data was not available, as is the case for J1250+3130. Clearly, the availability of {\it Spitzer} data provides much more information. 
}
\end{figure}

Taken at face-value, this indicates that the fraction of Type-1 QSOs at $z>6$ that are not detected in the WISE catalog are 
systematically and substantially fainter in WISE bands W1 and W2 than the WISE-detected QSOs; they are not just  
a little fainter, with the non-detections just evading detection below a 
formal flux limit. 
We now estimate the relative brightness of the WISE-detected and WISE-non-detected QSOs. We combine the results in bands W1 and W2, based on the results of the stacks described above. The average S/N ratio of the WISE-detected QSOs is 12.2 (the mean of 13.5 and 10.9), and the relative noise level in the stack of the non-WISE-detected QSOs is 1.26. Taking a 3-$\sigma$ limit for the non-WISE-detected sources, this gives a flux ratio of $12.2/(3 \times 1.26) \simeq 3.2$. We interpret this as meaning that the WISE non-detected QSOs in the sample considered here are likely to be on average a factor of about 3 times fainter than the WISE-detected QSOs. 

The number of galaxies in this stack is very modest, however, and so the statistical 
power is limited. Nevertheless, there does appear to be a real difference in the mid-IR flux densities of the WISE-detected and WISE-non-detected $z>6$ QSO  targets, by an amount  comparable to the offsets in their rest-UV optical magnitudes discussed further in Section\,3 below. 

\section{Discussion}

We now discuss the WISE properties of the $z>6$ QSOs in the context of their optical and near-IR emission. The components of the emission that contributes to the brighter WISE-detected 
examples are investigated using SED models in the WISE bands in Fig.\,4. 
They are not significantly different from those previously reported at low and modest redshifts for 
{\it Spitzer}-detected QSOs (Richards et al. 2006; Stern et al.\ 2012). Deep {\it Spitzer} observations of high-redshift 
QSOs (Jiang et al.\ 2010) have revealed some QSOs at $z \sim 6$ with little or no hot dust emission in the 24-$\mu$m band. 
The larger number of much shallower observations by WISE in the W3 (12-$\mu$m) band, which is sensitive to the 
hottest dust, could eventually be stacked to help to reveal whether these objects are unusual or typical. 
At face-value, our 
$z>6$ sample indicates no obvious difference in the shapes of the SEDs 
of the WISE-detected and WISE-non-detected $z>6$ QSOs considered here to the depth of the survey. 

There seems to be little possibility that the WISE colors could be affected dramatically by redshifted line emission. 
In the range $6<z<7$ Lyman-$\alpha$ lies in the $i$ and $z$ bands, while H$\alpha$ is in WISE W2, and H$\beta$ is in WISE W1. 
The average restframe equivalent width is 194\,\AA\, and 46\,\AA\, for H$\alpha$ and H$\beta$ from the SDSS QSO template of 
vanden Berk et al. (2001). Redshifting these lines from $z=6.5$ would correspond to equivalent widths of 0.15\,$\mu$m for 
H$\alpha$ in W2 and 0.035\,$\mu$m for H$\beta$ in W1 respectively. The fractional bandwidths are approximately 25\%, and so about 15\% of the W2 flux and 5\% of the W1 flux might be expected to be contributed by these lines. 
While there is evidence that the equivalent width of H$\alpha$ increases in galaxies at 
higher redshifts (Shim et al. 2011), this line contamination effect is unlikely to be significant here, in light of the size of the measurement uncertainties. 

\begin{figure}
\epsscale{1.15}
\plottwo{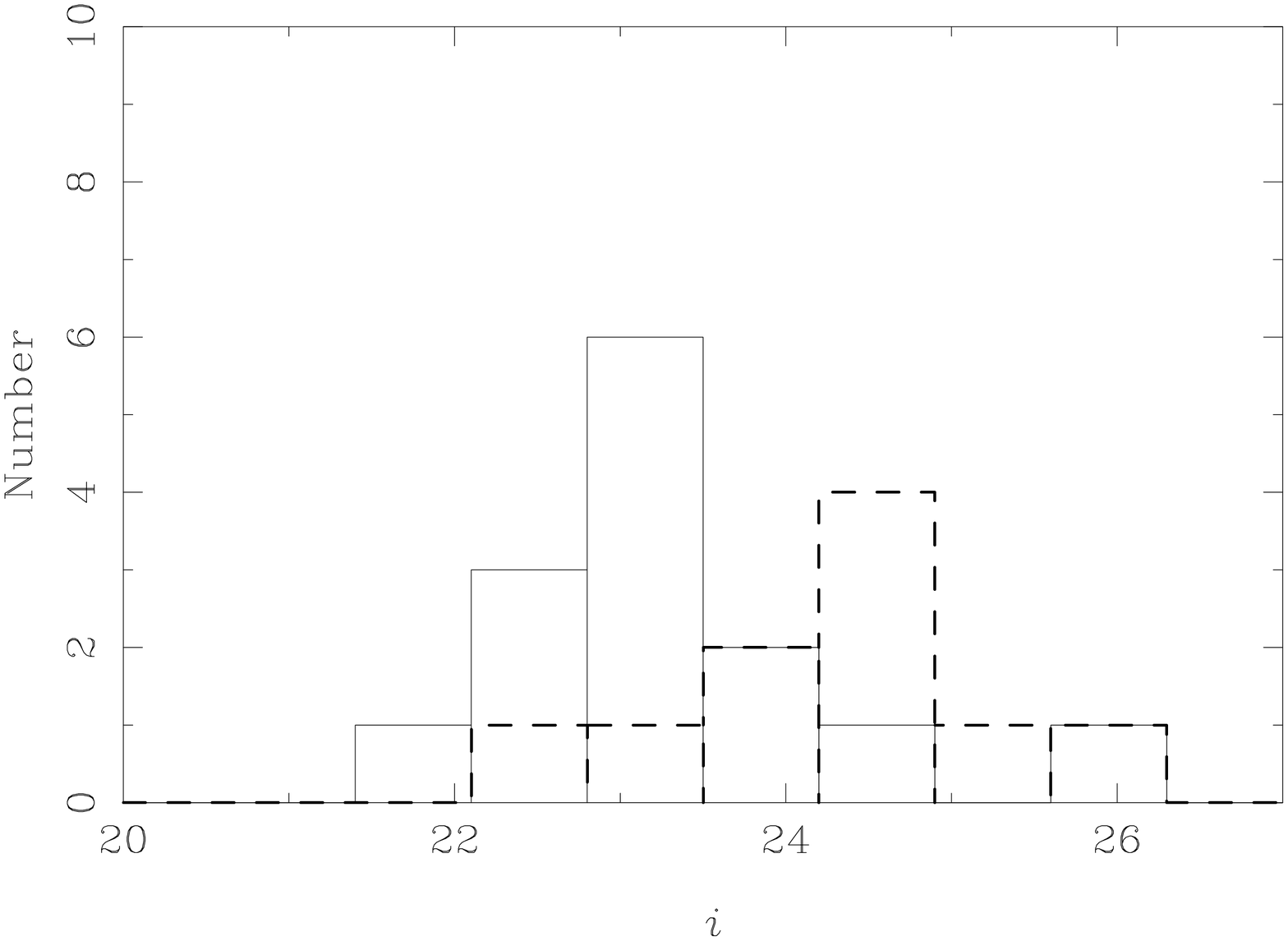}{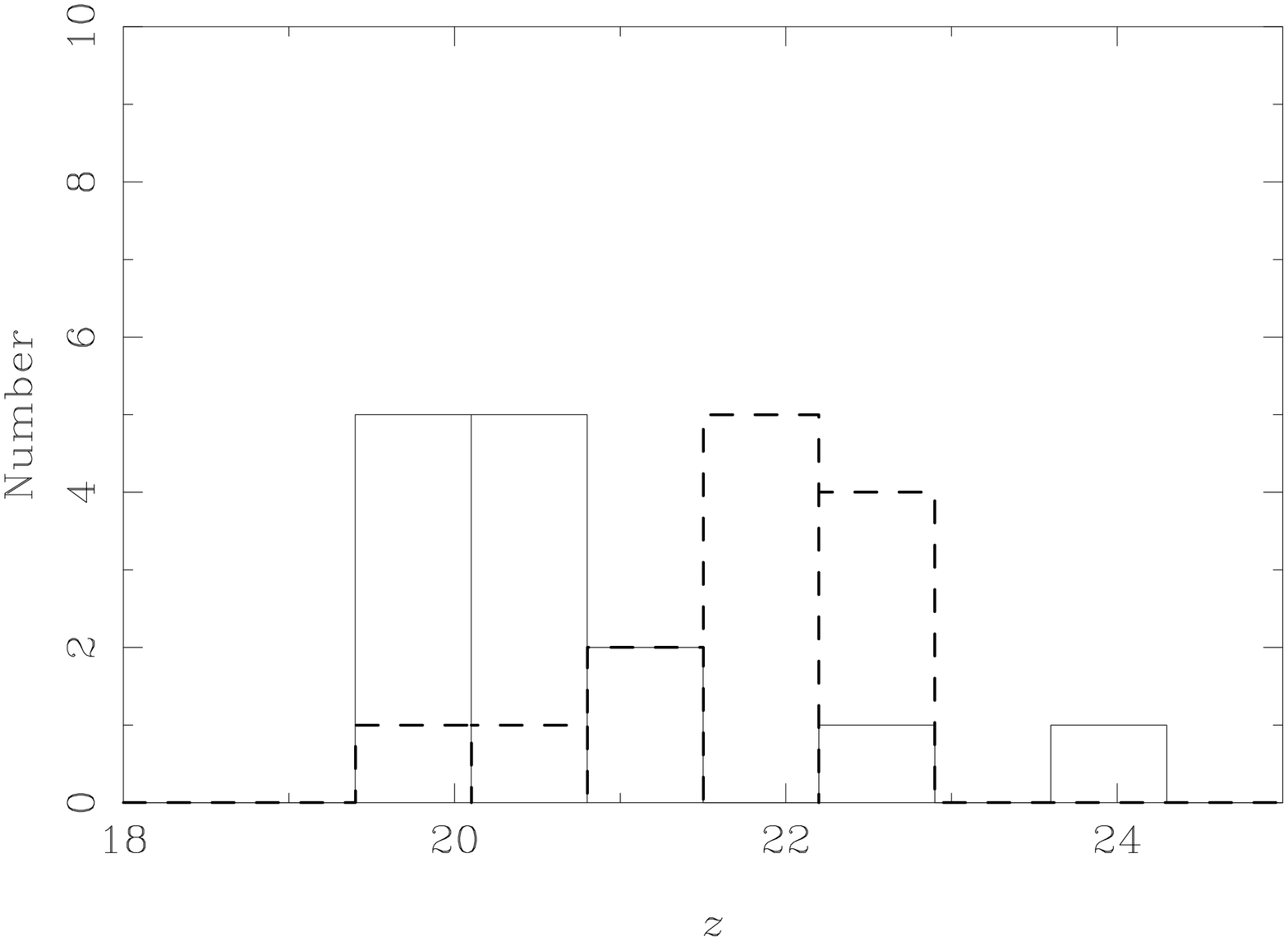}
\plottwo{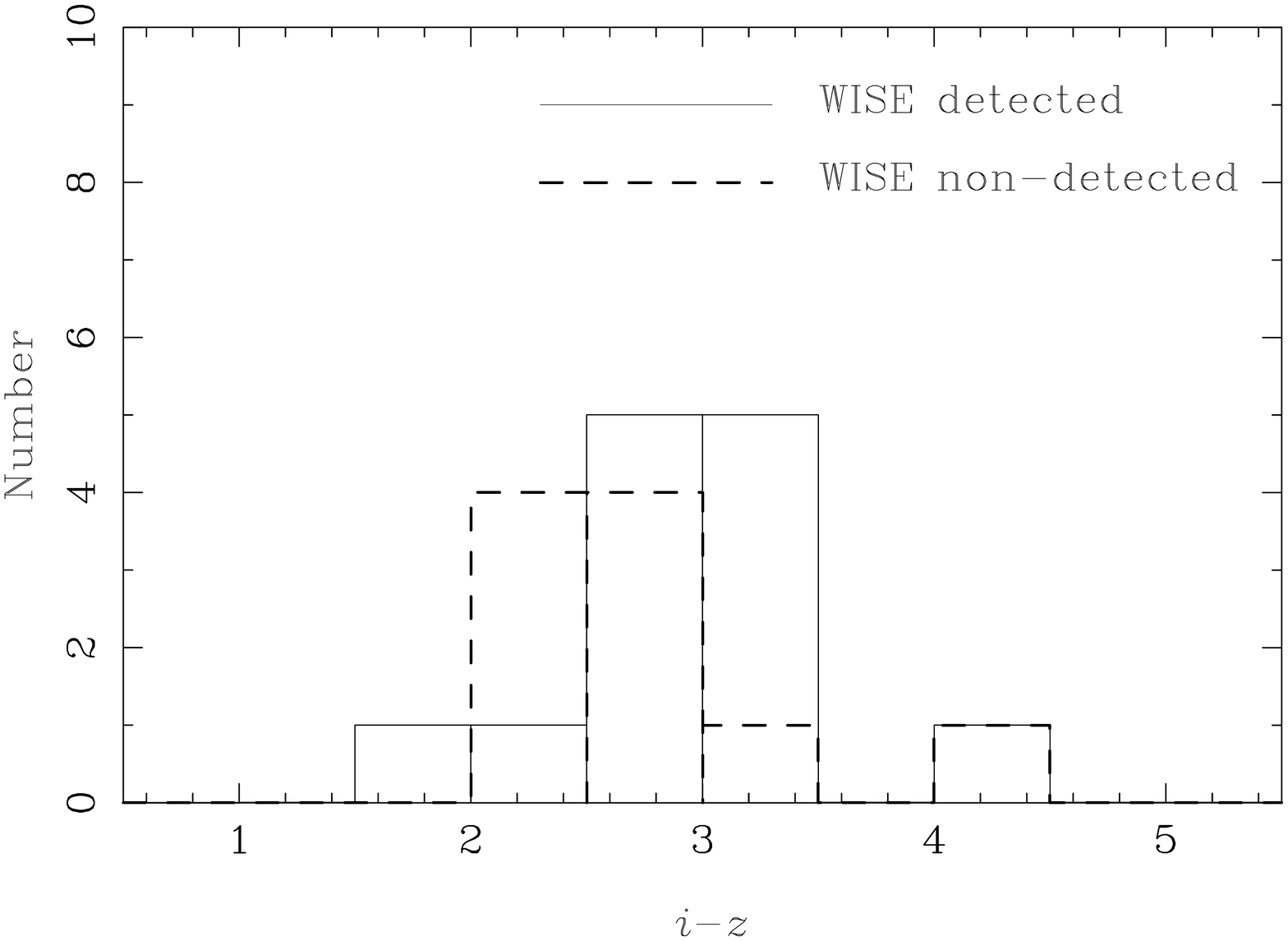}{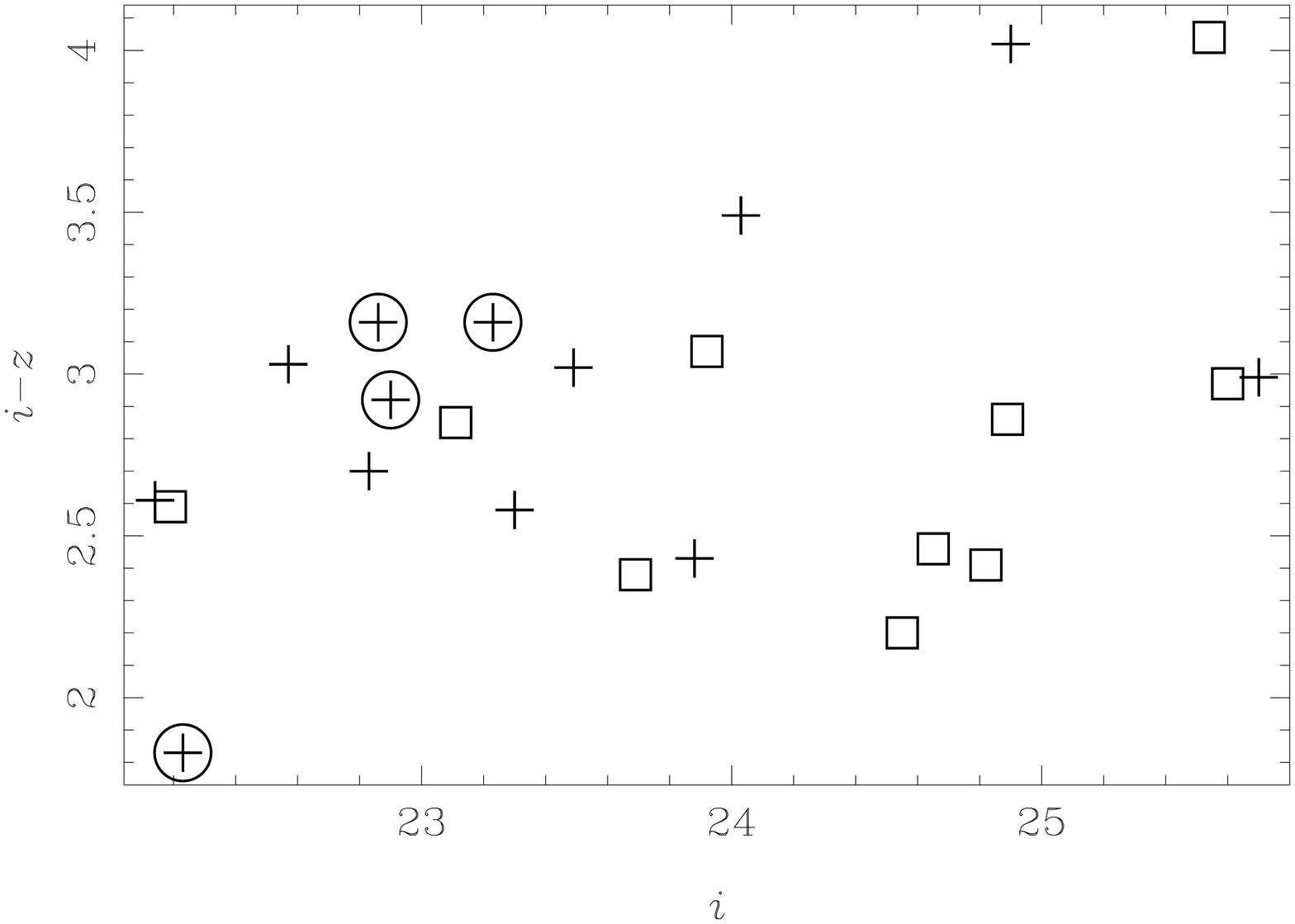}
\caption{Observed-frame optical and IR properties of $z>6$ QSOs with $i$- and/or $z$-band data. 
In the histograms, the WISE-detected (solid lines) and WISE-non-detected (dashed lines) QSOs have 
distinct properties, with the WISE-detected QSOs being brighter and redder at restframe 0.15\,$\mu$m. 
The color-magnitude diagram shows all 24 of the $z>6$ QSOs with $i$- and $z$-band data. 
Crosses show the 14 WISE-detected QSOs listed in Table\,1. Boxes represent WISE non-detected QSOs 
with $i$- and $z$-band data (10 in total). Circles highlight the QSOs with detections with {\it Spitzer} 
(Hines et al. 2006; Jiang et al. 2006; McGreer et al. 2006; Stern et al. 2007; Jiang et al. 2010, see Table\,1). 
An outlier, QSO\,J1630+4012, has been removed (Kim et al. 2009). All magnitudes are AB. 
}
\end{figure}


The WISE colors of the $z>6$ detections are compared with the WISE colors of larger samples of lower-redshift QSOs 
in Fig.\,1. The WISE--SDSS cross-matched sample has a much more complete coverage of low-redshift objects than other surveys. The SDSS 
sample contains relatively few lower-redshift QSOs 
around (4.2, 0.4) in the region of WISE color space where the $z>6$ QSOs are found (Fig.\,1); 
it includes a larger, more-complete sample of QSOs with matches in the WISE catalog at $3<z<5$, and extends to 
a maximum redshift 
of $z=5.41$. At modest redshifts, $z < 2-3$, the 
WISE colors provide a selection comparable to that described by Stern et al. (2005) for more moderate redshifts, 
due to the relative differences between the colors of evolved stellar light 
and the AGN power-law that dominates at longer wavelengths, as revealed by the W1$-$W2 color (see Stern et al. 2012, 
the discussion below and Fig.\,4). 
However, despite the 
confirmed ability of WISE to detect QSOs out to the end of the epoch of reionization, we emphasize that 
the colors of the WISE-detected 
objects do not stand out from the much larger numbers of AGN at more moderate 
redshifts (Fig.\,1), as their colors are determined by a broken power-law SED that does not change very 
strongly with increasing redshift. 
Thus the discovery of purely WISE-selected samples of high-redshift AGN are unlikely; there is no unambiguous 
signature of a very high-redshift QSO ($z>6$) from the WISE data alone at the WISE depth. The W1$-$W2 colors of the $z>6$ 
QSOs are typically bluer than those at low redshifts, as expected, but they are difficult to distinguish from $z \simeq 0.5$ 
star-forming galaxies as many are bluer than the ${\rm W1}-{\rm W2}>0.8$ color cut imposed in WISE-color-based AGN 
selection methods (see Stern et al. 2012; Mateos et al. 2012; Assef et al. 2013). 

Furthermore,  most of the W3 flux values are upper limits, and so the WISE points without error bars can move to the left in 
the W2-W3 color space of Fig.\,1, corresponding to bluer mid-IR colors. However, the detected $z>6$ QSOs remain within the cloud of lower-redshift QSOs and luminous IR galaxies (Wright et al.\ 2010; Jarrett et al. 2011) in this color space, 
preventing any easy sifting of QSO candidates from the WISE data. 

\begin{figure}
\epsscale{1.32}
\plotone{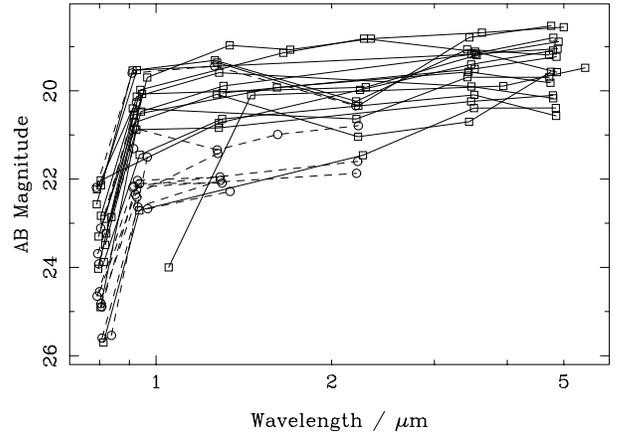}
\caption{The observed-frame optical--mid-IR SEDs of $z>6$ QSOs, from the $i$-band (0.8\,$\mu$m) to WISE W2 at 
4.6\,$\mu$m. The two rightmost clouds of points represent the WISE bands W1 and W2. The relative wavelength of 
different points in the clouds represent the relative redshift for each QSO, offset from the wavelength of the band 
center at a precise redshift  $z=6$.  
There are only three significant detections at longer wavelengths from WISE, and so we plot the SEDs only to 
restframe 5\,$\mu$m. 
WISE-detected QSOs have SEDs that extend longwards of 3\,$\mu$m on the plot, and 
are represented by square data points connected by solid lines, whereas WISE-non-detected QSOs are 
represented by circular data points detected by dashed lines. 
Note that WISE-detected examples have typically slightly redder observed 
$i-z$ colors (see Fig.\,5).
Note that at $z=6 (7)$, Lyman-$\alpha$ is at 0.85 (0.97)\,$\mu$m, potentially increasing the optical magnitudes, while H$\alpha$ potentially enhances the W2 flux according to the template in vanden Berk et al.\ (2001), as shown by the Assef et al. (2010) template curves in Fig.\,4. 
}
\end{figure}

WISE data add significantly to the available information about 
high-redshift candidates, as shown in Figs\,4, 5 \& 6. 
The SEDs of four of the brighter WISE detections, whose images are shown in Fig.\,2, are compared with templates 
from Assef et al. (2010) in Fig.\,4. These fits allow an estimate of the relative SED contributions 
of an evolved 
stellar component, and an AGN component. 
The AGN is described approximately by two power laws, with a break near 1\,$\mu$m (e.g. vanden Berk et al.\ 2001).  
These templates appear to provide a reasonable description of the SEDs of the
$z>6$ QSOs. In no cases is there compelling evidence that a large fraction of light from a young 
star-forming population (Sbc/Im) is required in the models to account for the data. Less precise SED information for more of the WISE-detected and non-detected QSOs 
is shown in Fig.\,6. 

Seven of the 17 WISE-detected $z>6$ QSOs listed in Table\,1 have {\it Spitzer} detections (see the footnotes to Table\,1), and 
they typically have a median of 3 bands of $izJHK$ photometry available, 
while the 14 non-WISE-detected $z>6$ QSOs have a median of 2 detections in these bands. 

The relative rest-UV colors of the WISE-detected and WISE-non-detected $z>6$ QSOs are compared in 
Fig.\,5. 
The 
relative magnitudes and colors, indicating that the WISE non-detections are both fainter, by about two magnitudes in $i$ and 
$z$, have a modestly greater range in their $z$-band magnitudes, and are marginally bluer in $i-z$ by about half a 
magnitude. As discussed above, $i-z$ straddles the wavelength of Lyman-$\alpha$ at these redshifts, and so this rest-UV color could be affected strongly by the column density of external and internal hydrogen absorption and the effects of 
extinction; 
note that the highest-redshift example, ULAS\,J1120+0641, does not appear in this plot, as Lyman-$\alpha$ has redshifted through both of the relevant bands at $z>7$. These differences are consistent with the WISE detectability amongst the 31 $z>6$ QSOs being related to the fainter $z$-band emission from the targets not detected by WISE. If the rest-UV light comes from accretion (Fig.\,4), and is on average 
two magnitudes fainter in the WISE-non-detected sample than the WISE-detected one, 
then it could just reflect significantly different accretion powers between the subsamples, 
without any major difference in the optical--mid-infrared colors. 
Note that there is inadequate deep near-IR data (see also Fig.\,6) to construct comparable histograms for the observed near-IR colors/magnitudes for these samples; some information is available from deeper observations (e.g.\ Staguhn et al. 2005), and all the available information is included in the SEDs shown in Fig.\,6. 

The large numbers of AGN candidates that are sure to be detected using PanSTARRS (Morganson et al.\ 2012), 
SkyMapper, VISTA, VST, DES and LSST can be matched to the WISE catalog 
in order to identify particularly dusty 
examples, to get a better idea of their true luminosity, and, at the highest redshifts, to obtain more useful limits on 
their stellar mass. The significant detection rate of existing $z>6$ QSO samples in the WISE All-Sky Release catalog, 
mostly from within the SDSS footprint, and in the deeper smaller-area CFHT surveys, offers a potentially large 
sample of future candidates. 
The WISE 
detection of a substantial fraction of known $z>6$ QSOs, and one at  $z>7$, makes us optimistic candidates at 
even higher redshifts could be found and isolated by taking advantage of a combination of deep wide-field optical--near-IR surveys 
and the WISE catalog. When WISE detects a QSO, and a redshift is known, the WISE data also immediately provides limits to its hot dust emission from W3, to its rest-optical luminosity from W1 \& W2, 
and even perhaps an evolved stellar population, all of 
which would otherwise be difficult to obtain until the launch of {\it JWST}.  
\section{Conclusions} 

We have searched for 31 known $z>6$ QSOs in the WISE All-Sky survey, and detected 17/31 (55\%) in band W1, and 
16, 3 and zero in bands W2, W3 and W4, respectively. This significant detection rate, for QSOs selected from both SDSS and 
deeper imaging, should make WISE a useful resource for exploiting ongoing wide optical/near-IR imaging surveys.  
The SED properties of the brighter WISE-detected objects are not unlike those of lower-redshift samples. 
From stacking the non-detected sample of 14 QSOs, their  3.4 and 4.6\,$\mu$m emission (in bands W1 and W2 
respectively) appears to be significantly less powerful than that of the WISE-detected QSOs, indicating that there is likely to be 
a significant difference between the mid-IR luminosities of the two classes, with WISE non-detection due to intrinsic 
faintness and not simply due to a matter of 
a chance appearance above the noise level in observations with low signal-to-noise ratios. 

\acknowledgments

We would like to thank an anonymous referee for their careful reading and helpful suggestions. 
This publication makes use of data products from the Wide-field Infrared Survey Explorer (WISE).
WISE is a joint project of the University of California, Los Angeles, and the Jet Propulsion Laboratory/California Institute of Technology, funded by the National Aeronautics and Space Administration.
This research has made use of both the NASA/IPAC Infrared Science Archive (IRSA) and the NASA/IPAC Extragalactic Database (NED), which are operated by the Jet Propulsion Laboratory, California Institute of Technology, under contract with the National Aeronautics and Space Administration. RJA was supported by an appointment to the NASA Postdoctoral
 Program at the Jet Propulsion Laboratory, administered by Oak Ridge 
 Associated Universities through a contract with NASA. 




{\it Facilities:} \facility{WISE}

\clearpage 

\begin{deluxetable}{lcclccccccccc}
\tabletypesize{\scriptsize}
\tablewidth{7.0in}
\tablecaption{
\small{
WISE Vega magnitudes for WISE-detected QSOs at $z>6$, where a WISE W1 signal-to-noise (S/N)
ratio is reported to be greater than 3.5. Error values are only listed for WISE cataloged results with 
reported S/N$>3.5$; otherwise, null results (...) are listed. The first fourteen entries are in the main WISE All-Sky
Release catalog, and the final three below the line are in the Reject Table. 
The offset between the reported QSO 
position and the WISE position is less than 1\,arcsec, apart from for 0227-0605, for which it is 1.46\,arcsec. 
In some bands the pipeline does not return a magnitude estimate, owing to high noise, or in a few cases, 
a nearby source. The zero points and the appropriate corrections for different spectral 
shapes are described by Wright et al. (2010) and Jarrett et al. (2011). 
}}
\tablewidth{0pt}
\tablehead{
\colhead{Name} & \colhead{RA} & \colhead{Dec} & Reference & $z$ & W1 & $\sigma_1$ 
& W2 & $\sigma_2$ & W3 & $\sigma_3$ & W4 & $\sigma_4$ 
}
\startdata
CFHQS J0050+3445 & 00:50:06.67 & +34:45:22.6 & Willott10a & 6.25 & 16.482 & 0.073 & 15.734 & 0.119 & 
13.084 & ... & 9.609 & ... \\
CFHQS J0227-0605 & 02:27:43.33 & $-$06:05:31.4 & Willott10a & 6.20 &  17.713 & 0.211 & 17.071 & ... & 
13.038 & ... & 9.498 & ... \\
SDSS J0353+0104$^a$ & 03:53:49.72 & +01:04:04.4 & Jiang08 & 6.049 & 16.886 & 0.147 & 16.497 & 0.375 & 
12.225 & ... & 8.560 & ... \\
SDSS J1030+0524$^{a,e}$ & 10:30:27.10 & +05:24:55.0 & Fan01 & 6.28 & 16.512 & 0.114  & 15.567 & 0.174 & 
12.364 & ... & 8.372 & ... \\
QSO J1048+4637$^{a,b,c,e}$ & 10:48:45.05 & +46:37:18.3 & Fan03 & 6.23 & 16.430 & 0.080 & 16.259 & 0.233 & 
12.885 & 0.495 & 8.843 & ... \\
ULAS J1120+0641 & 11:20:01.48 & +06:41:24.3 & Mortlock11 & 7.085 & 17.208 & 0.205 & 16.160 & 0.281 & 
12.523 & ... & 8.597 & ... \\
QSO J1137+3549$^{a}$ & 11:37:17.73 & +35:49:56.9 & Fan06 & 6.01 & 16.379 & 0.092 & 15.868 & 0.193 & 
12.144 & ... & 8.765 & ... \\
SDSS J1148+5251$^{b,d,e}$ &11:48:16.64 & +52:51:50.30 & Fan03 & 6.43 & 16.007 & 0.062 & 15.242 & 0.093 & 
12.544 & 0.352 & 8.598 & ... \\
QSO J1250+3130$^{a}$ & 12:50:51.93 & +31:30:21.9 & Fan06 & 6.13 & 16.489 & 0.096 & 15.474 & 0.136 & 
12.302 & ... & 8.473 & ... \\
ULAS J1319+0950$^b$ &13:19:11.29 &  +09:50:51.4 &  Mortlock09 &  6.127 & 17.222 & 0.145 & 16.848 & 0.400 & 
12.966  & ... & 9.030 & ... \\
QSO B1425+3326$^{a,h,j}$ & 14:27:38.59 & +33:12:42.0 & McGreer06 & 6.12 & 16.720 & 0.084 & 15.770 & 0.117 & 
12.801 & ... & 9.604 & ... \\
CFHQS J1429+5447 & 14:29:52.17 & +54:47:17.7 & Willott10a & 6.21 &  17.405 & 0.141 & 17.246 & 0.435 & 
12.903 & ... & 9.564 & ... \\
QSO J1602+4228$^{a,e}$&  16:02:53.98 &  +42:28:24.9 & Fan04 & 6.07 & 16.107 & 0.046 & 15.209 & 0.062 & 
12.184 & 0.158 & 9.526 & ... \\
QSO J1623+3112$^{a,e,f}$ & 16:23:31.81 & +31:12:00.50 & Fan04 & 6.22 & 16.839 & 0.110 & 15.914 & 0.166 & 
12.706 & ... & 9.268 & ... \\
\hline
QSO J1630+4012$^{a,e,i}$ & 16:30:33.90 & +40:12:09.60 & Fan03 & 6.05 & 18.000 & 0.273 & 17.138 & ... & 
13.122 & ... & 9.447 & ... \\
SDSS J2054-0005 & 20:54:06.49 & $-$00:05:14.80 & Jiang08 & 6.062 & 18.017 & 0.337 & 16.25 & ... & 
12.595 & ... & 8.727 & ... \\
SDSS J2315-0023 & 23:15:46.57 & $-$00:23:58.1 & Jiang08 & 6.117 & 17.559 & 0.283 & 16.785 & ... & 12.606 & ... & 8.624 & ... \\
\enddata
\tablewidth{7.0in}
\tablenotetext{a}{These QSOs were observed but not detected in the mm-wave continuum (Wang et al. 2007, 2008b).}
\tablenotetext{b}{These QSOs were detected in the mm-wave continuum (Wang et al. 2011b), implying a very large far-IR luminosity.}
\tablenotetext{c}{1048+4637 was detected by at submm wavelengths by Wang et al. (2008a), implying a very large far-IR luminosity.}
\tablenotetext{d}{1148+5251 was detected by {\it Spitzer} (Hines et al. 2006; Jiang et al. 2006), with consistent WISE magnitudes. 
It was also detected in the continuum at submm wavelengths by Beelen et al. (2006) 
and using {\it Herschel} by Leipski et al. (2010), implying a very large far-IR luminosity. A [CII] far-IR emission line was 
measured by Maiolino et al. (2012).}
\tablenotetext{e}{These galaxies were detected by {\it Spitzer} (Jiang et al. 2006),  with consistent magnitudes to WISE.}
\tablenotetext{f}{An upper limit to a CO line flux was reported for this QSO by Wang et al. (2011a).}
\tablenotetext{g}{Observed at 450 and 850\,$\mu$m by Robson et al. (2004). J1048+4637 was detected.}
\tablenotetext{h}{Detected by {\it Spitzer} (Stern et al. 2007).}
\tablenotetext{i}{An upper limit to CO line flux was reported for this galaxy by Maiolino et al. (2007).}
\tablenotetext{j}{QSO B1425+3326 is FIRST J1427385+331241 (McGreer et al. 2006), IRAC J142738.5+331242 (Stern et al.\ 2007) and SDSS J142738.59+331242.0 (Schneider et al.\ 2007).}
\end{deluxetable}

\end{document}